\documentclass{article}
\usepackage{spconf,amsmath,graphicx}
\usepackage{amsfonts} 
\usepackage{placeins}

\usepackage{graphicx}
\usepackage{float}

\usepackage{tikz}

\newcommand\submittedtext{%
  \footnotesize 
  Accepted at IEEE International Symposium on Biomedical Imaging 2020.\\© 2023 IEEE.  Personal use of this material is permitted.  Permission from IEEE must be obtained for all other uses, in any current or future media, including reprinting/republishing this material for advertising or promotional purposes, creating new collective works, for resale or redistribution to servers or lists, or reuse of any copyrighted component of this work in other works.}

\newcommand\submittednotice{%
\begin{tikzpicture}[remember picture,overlay]
\node[anchor=south,yshift=10pt] at (current page.south) {\fbox{\parbox{\dimexpr0.65\textwidth-\fboxsep-\fboxrule\relax}{\submittedtext}}};
\end{tikzpicture}%
}

\title{Transcriptome-supervised classification of tissue morphology using deep learning}
\name{Axel Andersson, Gabriele Partel, Leslie Solorzano, Carolina W\"ahlby \thanks{Thanks to the European  Research Council, grant  ERC‐2015‐CoG  682810 to Carolina W\"ahlby for funding and the Nilsson lab for providing image data.}}
\address{Dept. of Information Technology and SciLifeLab, Uppsala University,  Sweden}

\begin{document}
%
\maketitle
\submittednotice

\begin{abstract}
	Deep learning has proven to successfully learn variations in tissue and cell morphology. 
	Training of such models typically relies on  expensive manual annotations. Here we conjecture that spatially resolved gene expression, e.i., the transcriptome, can be used as an alternative to manual annotations. In particular, we trained five convolutional neural networks with patches of different size extracted from locations defined by spatially resolved gene expression. The network is trained to classify tissue morphology related to two different genes, general tissue, as well as background, on an image of fluorescence stained nuclei in a mouse brain coronal section. Performance is evaluated on an independent tissue section from a different mouse brain, reaching an average Dice score of 0.51. Results may indicate that novel techniques for spatially resolved transcriptomics together with deep learning may provide a unique and unbiased way to find genotype-phenotype relationships. 
	
\end{abstract}

\begin{keywords}
	In situ sequencing, Gene expression, Tissue classification, Deep learning
\end{keywords}
\section{Introduction}
\label{sec:intro}
The spatial organisation of different cell types provides us with cues on tissue function, and allows us to study complex organs like the brain in relation to development or disease. Deep learning has proven to successfully learn variations in tissue and cell morphology, and thereby providing objective solutions for cancer diagnosis and grading based on the content of microscopy images~\cite{strom2020artificial, BreastCancerGrading1, cellularmorphwithmanualannotations}. Training of such models typically relies on data manually labelled by expert pathologists. Although networks trained on manual labels  have shown promising results, manual labelling is tedious, subjective, and prone to inter-observer variations~\cite{InterObserverVariability, InterObserverVariability2}.
Such labelling will also rely on a priory known patterns. 

An alternative to manual labelling is to physically add fluorescent labels that binds to particular cellular constituents. Unfortunately, this type of labels may perturb the morphology of the experimental sample. Spectral overlap also limits the total number of fluorescent labels, making this approach impractical for deep learning purposes. Christiansen et al.~\cite{christiansen2018silico} and Ounkomol et al.~\cite{ounkomol2018label} used deep learning to predict fluorescent labels directly from transmitted-light images, allowing fluorescent labels without physically perturbing the experimental samples. This so called in silico labelling is however not designed for defining variations in morphology related to disease.

Meanwhile, modern molecular detection techniques, like \textit{in situ} sequencing (ISS) ~\cite{Ke2013}, enables detection of gene expression directly in biological tissue samples without loss of spatial morphology. The highly multiplex output from ISS also allows for automated labelling of multiple cell types~\cite{Qian2018} as well as labelling of several tissue compartments~\cite{Partel2019}; fully automated annotations that otherwise would be difficult to obtain.

Here we explore the idea of using spatially resolved gene expressions, or transciptomics, obtained with ISS, as annotation for classification of tissue morphology. We train a convolutional neural network (CNN) with aim of classifying local tissue compartments by their morphology. Training patches are generated from  regions defined by the gene expression, from an image of a nuclei fluorescent stained mouse brain coronal section.

\section{Material and Methods}
\subsection{Image Data}
Our data consists of two highly resolved,  fluorescence microscopy images of mouse brain sections (collected in accordance with the directives and guidelines of the Swedish Board of Agriculture, the Swedish Animal Protection Agency, and the Karolinska Institutet). 

\begin{figure*}[ht]
	\centering
	
	\includegraphics[width=0.98\linewidth]{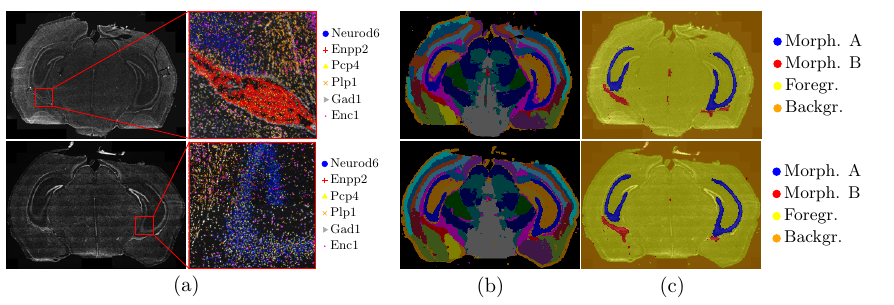}
	
	\caption{(a) Nuclei stained coronal sections of two mouse brains together with zoomed in regions with markers for gene expression. (b) Twenty different regions derived from the gene expression using dimensionality reduction and clustering presented in~\cite{Partel2019}. (c) We focus on two regions from (b) that are primary defined by a single gene each, namely Neurod6 and Enpp2. The morphology within these two regions are referred to as Morpholgoy A and Morphology B. We also introduce a mask for any other type of morphology (foreground), as well as a mask for the background. Patches are then to be cropped from within these regions and used for training the neural network. Top row corresponds to the train image, bottom row corresponds to the test image.}
	\label{fig:markers}
\end{figure*}

We will refer to them as the train image and the test image. We also have access to spatially resolved gene expression in form of 82 different ISS marker types, each of which corresponding to a particular gene~\cite{Ke2013}. The two images together with markers is shown shown Figure~\ref{fig:markers}~(a).

\subsection{Generating the Dataset}
Training a classifier typically requires annotated training data. In our case, we require images of tissue labelled by different morphology types. We hypothesise that cells with similar gene expression are likely to be of the same type and therefore attain similar morphological characteristics. Our approach is to delineate areas for different morphologies by first localising regions of similar gene expression. To automatically identify these regions of similar gene expression, we apply the dimensionality reduction and clustering described in~\cite{Partel2019}. The resulting regions obtained after the gene expression clustering are shown in Figure~\ref{fig:markers}~(b).

In~\cite{Partel2019}, regions can be defined by combinations of many genes, but in this preliminary study we focus on two regions primary defined by a single gene each, namely Neurod6 and Enpp2.  We will refer the morphology within these two regions as Morphology A and Morphology B. Patches labelled by either Morphology A or Morphology B are then cropped around markers for Neurod6 and Enpp2 from within these two regions.

Furthermore, we want our classifier to be able to distinguish between our two morphologies of interest and other types of morphologies. This motivates us to introduce two additional classes, namely a background class, corresponding to images not containing any nuclei, as well as a foreground class, corresponding to morphology that is neither Morphology A nor Morphology B.  
\begin{figure}[hb!]
    \centering
	\begin{minipage}[b]{0.18\linewidth}
		\centering
		\centerline{\includegraphics[width=\linewidth, height=\linewidth]{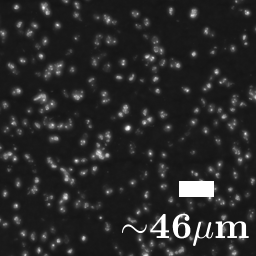}}
	\end{minipage}
	\begin{minipage}[b]{0.18\linewidth}
		\centering
		\centerline{\includegraphics[width=\linewidth, height=\linewidth]{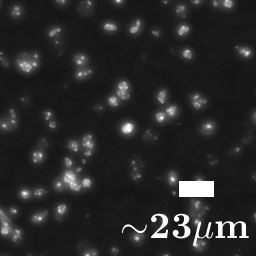}}
	\end{minipage}
	\begin{minipage}[b]{0.18\linewidth}
		\centering
		\centerline{\includegraphics[width=\linewidth, height=\linewidth]{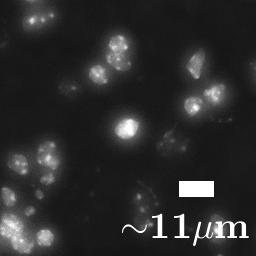}}
	\end{minipage}
	\begin{minipage}[b]{0.18\linewidth}
		\centering
		\centerline{\includegraphics[width=\linewidth, height=\linewidth]{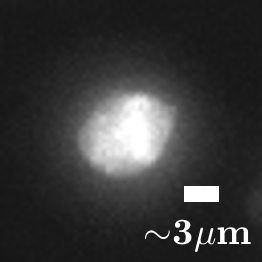}}
	\end{minipage}

	\begin{minipage}[b]{0.18\linewidth}
		\centering
		\centerline{\includegraphics[width=\linewidth, height=\linewidth]{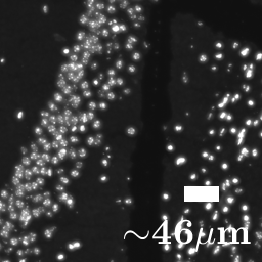}}
		\vspace{0.2cm}
		\centerline{\footnotesize{$1024\times1024$}}\medskip
	\end{minipage}
	\begin{minipage}[b]{0.18\linewidth}
		\centering
		\centerline{\includegraphics[width=\linewidth, height=\linewidth]{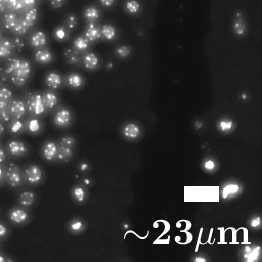}}
		\vspace{0.2cm}
		\centerline{\footnotesize{$512\times512$}}\medskip
	\end{minipage}
	\begin{minipage}[b]{0.18\linewidth}
		\centering
		\centerline{\includegraphics[width=\linewidth, height=\linewidth]{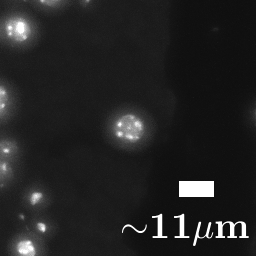}}
		\vspace{0.2cm}
		\centerline{\footnotesize{$256\times256$}}\medskip
	\end{minipage}
	\begin{minipage}[b]{0.18\linewidth}
		\centering
		\centerline{\includegraphics[width=\linewidth, height=\linewidth]{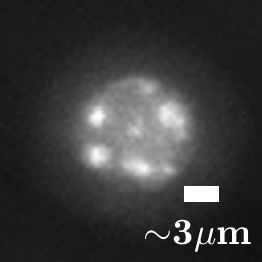}}
		\vspace{0.2cm}
		\centerline{\footnotesize{$64\times64$}}\medskip
	\end{minipage}
	\caption{Sample training patches of different resolutions.}
	\label{fig:resolutions}
\end{figure}

\begin{figure*}[htb!]
    \FloatBarrier
	\centering
	\begin{minipage}{0.97\linewidth}
		\centering
		\centerline{\includegraphics[width=\linewidth]{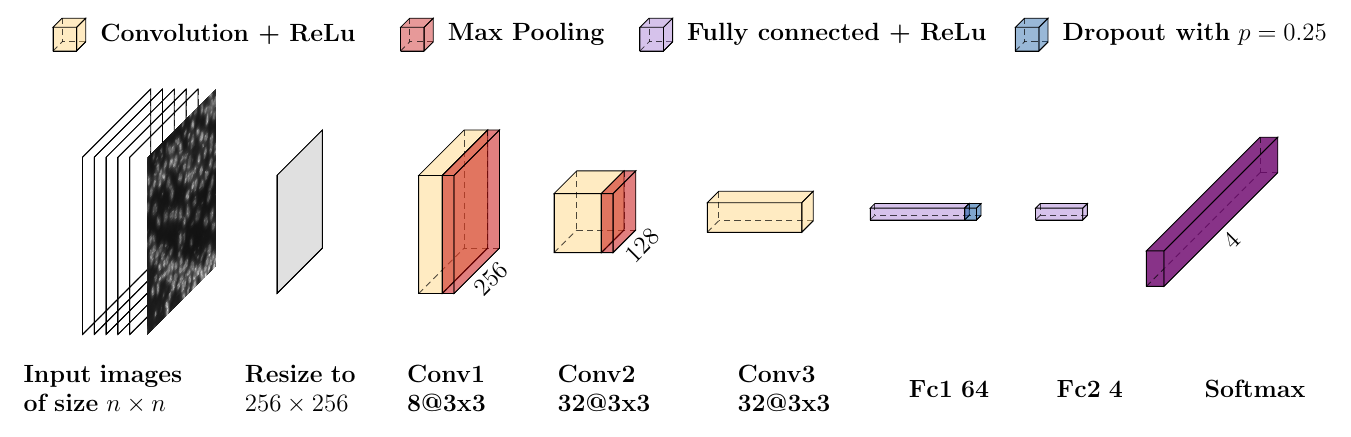}}
	\end{minipage}
	\caption{Network architecture}
	\label{fig:network}
\end{figure*}
To crop patches of the foreground, we first need to mask the foreground. This mask is obtained by taking the union of all regions that do correspond to neither Morphology A nor Morphology B. Similarly, a mask for the background is obtained by taking the compliment to the union of the foreground mask and the masks for Morphology A and Morphology B. Patches labelled as background and foreground are then cropped from random locations within these two masks. The four regions from which annotated patches are cropped is shown in Figure~\ref{fig:markers}~(c).

Intuitively, the size of the patches should also affect the performance of the classifier. A large patch-size would allow the network to not only pick up on characteristics of individual nuclei, but also take into account features associated with the surrounding tissue. However, a too large patch-size may contain morphology associated with other classes. To investigate the effect of the different patch sizes, we choose to generate five different data sets, where the width and height of the patches in each set are $64$, $128$, $256$, $512$ and $1024$ pixels. Figure~\ref{fig:resolutions} shows patches of different sizes.

\subsection{The network}
We created a CNN using the Deep Learning Toolbox in {\tt Matlab 2018b}. The model consists of three convolutional layers, each alternated by max pooling layers, followed by two fully connected layers and ending with a softmax layer, see Figure~\ref{fig:network}. A rectified linear unit function is used as activation function between each of the layers. The network is trained with ADAM optimiser~\cite{ADAM}. Images were resized to a size of $256 \times 256$ using bi-linear interpolation with $2\times 2$ nearest neighbours before being fed to the network. The training images were augmented with a random horizontal and vertical flip, as well as a random rotation sampled from the interval $[0,360]$.  The training lasts until the loss, calculated on the test set after every epoch, has not increased for two consecutive epochs. 
[htb!]

\subsection{Evaluation}
We evaluate our trained classifier by striding it across the fully resolved test image. If we slide the classifier with a stride of $1$ every pixel in the image gets associated with a class. Since the number of pixels in the test image is of order $10^8$, evaluating the classifier at every pixel location is too expensive. To speed up the evaluation, we use a horizontal and vertical stride of $32$ and then up-scale using  nearest neighbour interpolation. We then compare the similarity between the CNN predictions and the masks for Morpohology A, Morphology B, Background and Foreground, by computing the S{\o}rensen-Dice similarity coefficient (SDC). Mathematically, the SDC for the k'th class is computed as
\begin{equation*}
	SDC^{(k)} = \frac{2|{I}^{(k)} \cap \tilde{I}^{(k)}|}{|{I}^{(k)}| + |\tilde{I}^{(k)}|},
\end{equation*}
where $\tilde{I}^{(k)}$ is predicted segmentation of the k'th class, structured as a binary image and $I^{(k)}$ is the k'th binary mask.

\section{Results}
\begin{figure}[ht]
	\begin{minipage}[b]{1.0\linewidth}
		\centering
		\centerline{\includegraphics{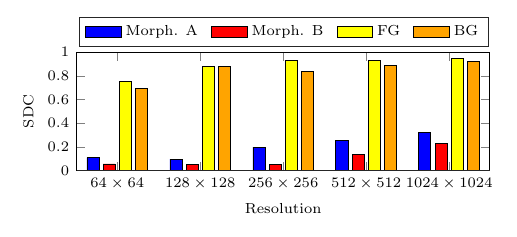}}
	\end{minipage}
	\caption{Dice score for different patch sizes.}
	\label{fig:dice}
\end{figure}

We trained the classifier on five separate datasets. The different datasets consist of images cropped at the same locations, but with different patch sizes. The SDC for each of the data sets is shown in Figure~\ref{fig:dice}.

\begin{figure}[ht]
	\centering
	
	\begin{minipage}[b]{0.46\linewidth}
		\centerline{\includegraphics[scale=.8]{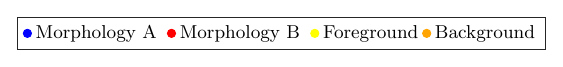}}
	\end{minipage}
	
	\begin{minipage}[b]{0.44\linewidth}
		\centerline{\includegraphics[width=3.8cm]{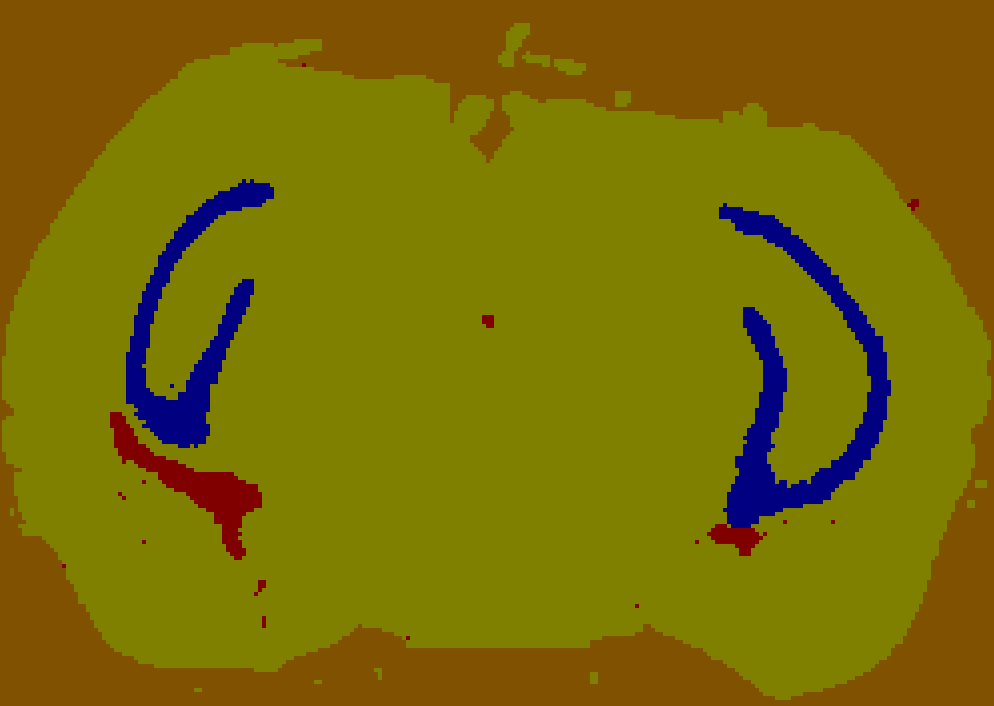}}
	\end{minipage}
	\begin{minipage}[b]{0.44\linewidth}
		\centerline{\includegraphics[width=3.8cm]{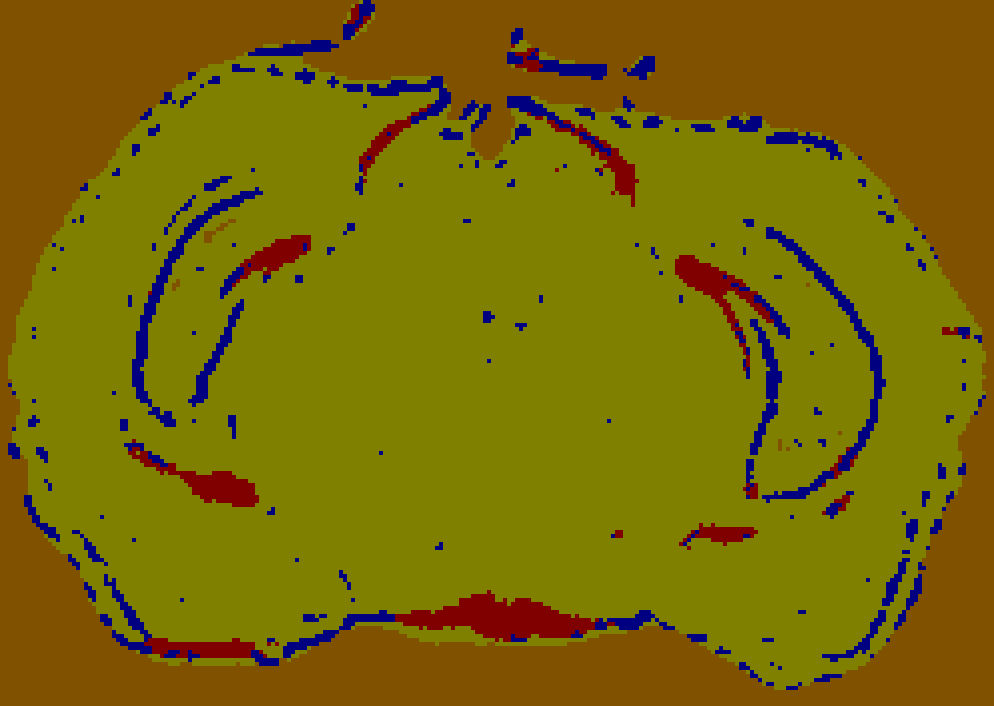}}
	\end{minipage}
	\caption{Ground truth (left) and prediction result (right).}
	\label{fig:result}
\end{figure}

Figure~\ref{fig:result} shows ground-truth masks together with the masks predicted by the network trained on patches of size $1024 \times 1024$.

\section{Discussion and conclusion}
\label{sec:discussion}
Judging by the SDC in Figure~\ref{fig:dice}, the network performs better when trained on larger patches. Hypothetically, it is easier for the network to classify the two different morphologies based on features associated with groupings of cells, such as cell density or distance between cells. Although, it is worth noting that the patches in the data sets are not centred around cell nuclei. This could potentially cause trouble when training on smaller patch sizes. For instance, if patches do not contain the entire body of a nucleus, it will be difficult for the network to pick up on features associated with the morphology of individual nuclei. 

Moreover, it appears visually from Figure~\ref{fig:result} as if the network has a tendency of miss-classifying regions of the Foreground as either Morphology A or B. Inspecting the morphology in these false-positive regions reveals a higher nuclei density, hence it is likely that nuclei density is one of the features that the network uses for classification. It is also interesting to see that some of the false-positive regions in Figure~\ref{fig:result} show a similar spatial pattern as the clusters of gene expressions shown in Figure~\ref{fig:markers}~(b).

It is also worth mentioning that the foreground class encapsulates many different types of morphologies. By splitting the foreground class into finer sub-classes, such that each sub-class corresponds to a distinct morphology, one could potentially improve the discriminative ability of the network. This will be explored in future work. Furthermore, we want to stress that there is very limited amount of morphological information when only staining for DNA in the nuclei. In future work, we  hope to continue exploring techniques for relating genotype with phenotype using data with richer morphological information, for instance, hematoxylin and eosin stained tissue slides.

\bibliographystyle{IEEEbib}
\bibliography{refs}

\begin{thebibliography}{10}

\bibitem{strom2020artificial}
Peter Str{\"o}m, Kimmo Kartasalo, Henrik Olsson, Leslie Solorzano, Brett
  Delahunt, Daniel~M Berney, David~G Bostwick, Andrew~J Evans, David~J Grignon,
  Peter~A Humphrey, et~al.,
\newblock ``Artificial intelligence for diagnosis and grading of prostate
  cancer in biopsies: a population-based, diagnostic study,''
\newblock {\em The Lancet Oncology}, 2020.

\bibitem{BreastCancerGrading1}
Haibo Wang, Angel Cruz-Roa, Ajay Basavanhally, Hannah Gilmore, Natalie Shih,
  Mike Feldman, John Tomaszewski, Fabio Gonzalez, and Anant Madabhushi,
\newblock ``Mitosis detection in breast cancer pathology images by combining
  handcrafted and convolutional neural network features,''
\newblock {\em J. of Medical Imaging}, vol. 1, no. 3, pp. 034003, Oct. 2014.

\bibitem{cellularmorphwithmanualannotations}
Lenka Strbkova, Daniel Zicha, Pavel Vesely, and Radim Chmelik,
\newblock ``Automated classification of cell morphology by coherence-controlled
  holographic microscopy,''
\newblock {\em J. of biomedical optics}, vol. 22, no. 8, pp. 086008, 2017.

\bibitem{InterObserverVariability}
Douglas~S Gomes, Simone~S Porto, D{\'{e}}bora Balabram, and Helenice Gobbi,
\newblock ``Inter-observer variability between general pathologists and a
  specialist in breast pathology in the diagnosis of lobular neoplasia,
  columnar cell lesions, atypical ductal hyperplasia and ductal carcinoma in
  situ of the breast,''
\newblock {\em Diagnostic Pathology}, vol. 9, no. 1, pp. 121, 2014.

\bibitem{InterObserverVariability2}
Thomas~J. Lawton, Geza Acs, Pedram Argani, Gelareh Farshid, Michael Gilcrease,
  Neal Goldstein, Frederick Koerner, J.~Jordi Rowe, Melinda Sanders, Sejal~S.
  Shah, and Carol Reynolds,
\newblock ``Interobserver variability by pathologists in the distinction
  between cellular fibroadenomas and phyllodes tumors,''
\newblock {\em Int. J. of Surgical Pathology}, vol. 22, no. 8, pp. 695--698,
  Aug. 2014.

\bibitem{christiansen2018silico}
Eric~M Christiansen, Samuel~J Yang, D~Michael Ando, Ashkan Javaherian, Gaia
  Skibinski, Scott Lipnick, Elliot Mount, Alison O’Neil, Kevan Shah, Alicia~K
  Lee, et~al.,
\newblock ``In silico labeling: predicting fluorescent labels in unlabeled
  images,''
\newblock {\em Cell}, vol. 173, no. 3, pp. 792--803, 2018.

\bibitem{ounkomol2018label}
Chawin Ounkomol, Sharmishtaa Seshamani, Mary~M Maleckar, Forrest Collman, and
  Gregory~R Johnson,
\newblock ``Label-free prediction of three-dimensional fluorescence images from
  transmitted-light microscopy,''
\newblock {\em Nature methods}, vol. 15, no. 11, pp. 917, 2018.

\bibitem{Ke2013}
Rongqin Ke, Marco Mignardi, Alexandra Pacureanu, Jessica Svedlund, Johan
  Botling, Carolina W\"{a}hlby, and Mats Nilsson,
\newblock ``In situ sequencing for {RNA} analysis in preserved tissue and
  cells,''
\newblock {\em Nature Methods}, vol. 10, no. 9, pp. 857--860, July 2013.

\bibitem{Qian2018}
Xiaoyan Qian, Kenneth~D Harris, Thomas Hauling, Dimitris Nicoloutsopoulos,
  Ana~Munoz Manchado, Nathan Skene, Jens Hjerling-Leffler, and Mats Nilsson,
\newblock ``A spatial atlas of inhibitory cell types in mouse hippocampus,''
\newblock {\em bioRxiv preprint doi:10.1101/431957}, 2018.

\bibitem{Partel2019}
Gabriele Partel, Markus~M. Hilscher, Giorgia Milli, Leslie Solorzano, Anna~H.
  Klemm, Mats Nilsson, and Carolina W{\"a}hlby,
\newblock ``Identification of spatial compartments in tissue from in situ
  sequencing data,''
\newblock {\em bioRxiv preprint doi:10.1101/765842}, 2019.

\bibitem{ADAM}
Diederik~P Kingma and Jimmy Ba,
\newblock ``Adam: A method for stochastic optimization,''
\newblock {\em arXiv preprint arXiv:1412.6980}, 2014.

\end{thebibliography}

\end{document}